# Interfacing whispering-gallery microresonators and free space light with cavity enhanced Rayleigh scattering

Jiangang Zhu[1*], Şahin K. Özdemir[1*], Huzeyfe Yilmaz[1], Bo Peng[1], Mark Dong[2], Matthew Tomes[2], Tal Carmon[2], and Lan Yang[1*]

[1]Department of Electrical and Systems Engineering, Washington University, St. Louis, MO 63130, USA

[2]Department of Electrical and Computer Engineering, University of Michigan, Ann Arbor, MI 48109, USA

*Correspondence to: yang@ese.wustl.edu, ozdemir@ese.wustl.edu, jzhu@seas.wustl.edu

**Whispering gallery mode resonators (WGMRs) take advantage of strong light confinement and long photon lifetime for applications in sensing, optomechanics, microlasers and quantum optics. However, their rotational symmetry and low radiation loss impede energy exchange between WGMs and the surrounding. As a result, free-space coupling of light into and from WGMRs is very challenging. In previous schemes, resonators are intentionally deformed to break circular symmetry to enable free-space coupling of carefully aligned focused light, which comes with bulky size and alignment issue that hinder the realization of compact WGMR applications. Here, we report a new class of nanocouplers based on cavity enhanced Rayleigh scattering from nano-scatterer(s) on resonator surface, and demonstrate whispering gallery microlaser by free-space optical pumping of an Ytterbium doped silica microtoroid via the scatterers. This new scheme will not only expand the range of applications enabled by WGMRs, but also provide a possible route to integrate them into solar powered green photonics.**

The last two decades have witnessed a revolution in photonic technologies pioneered on one hand by new concepts in materials and devices such as photonic crystals and meta materials, and, on the other hand, by the realization and testing of century-old well-known theories such as quantum theory, plasmonics and whispering galleries which have been enjoying many benefits of recent developments in enabling technologies and fabrication techniques[1-5]. Since its first explanation in acoustic regime by Lord Rayleigh in London's St Paul's Cathedral, WGM phenomenon has been explored in various optical structures for a variety of applications[6,7], opening unprecedented and unforeseen directions in optical sciences. Recent advances in fabrication techniques and material sciences have helped to achieve WGMRs with ultra-high-quality (Q) factors and nano/micro-scale mode volumes(V), which in turn have enabled novel applications and devices such as ultra-low threshold on-chip microlasers[8-11], narrowband filters and modulators for optical communication[12-14], high performance optical sensors achieving label-free detection at single-particle resolution[1,4,5,15,16], cavity opto-mechanics[2,17,18] and quantum electrodynamics[3,19].

Despite their great promises for photonic technologies, coupling light into and from WGMRs is intrinsically hindered by their unique feature of rotational symmetry. The circular geometry is also responsible for the deviation from total internal reflection condition, and introduces radiation losses in particular when the wavelength of the light is comparable to the radius of curvature. Thus, an evanescent field channel exists for extracting light from WGMs, and for coupling light into WGMs using prisms, tapered fibers or waveguides[7,20-22]. Coupling efficiency with tapered fibers can reach values as high as 99%. However, achieving this coupling and maintaining it for long durations require active stabilization and precise alignment with nanopositioning systems, because coupling conditions are prone to environmental perturbations (e.g., air flow and mechanical vibrations). This significantly limits the practical use of fiber-taper-coupled WGMRs. Alternative to evanescent coupling techniques is fabricating asymmetric WGMRs such as spiral, stadium, ellipsoid, quadrupole and limaco[23-25]. There are also studies with well-known symmetric WGMRs, such as microspheres, microdisks and microtoroids, where circular symmetry is lifted by introducing controlled deformations either after the WGMRs are fabricated or during lithographic patterning[26]. Free-space coupling into and directional emission from deformed/asymmetric resonators are possible due to the dynamic tunneling between the co-existing chaotic and regular WGM modes, which help the light to escape from or couple into the resonator along the direction of deformation. Coupling of free space light into such resonators still remains as a challenge, mostly because it relies significantly on precise alignment of the focused free-space light on the cavity edge along the direction of deformation, which requires optical and mechanical systems with high angular and translational resolution. These unavoidably make the system bulky and difficult to move out of the lab environment. Moreover, with the exception of a few studies, such cavities suffer from significant Q-degradation as the degree of deformation is increased. Here we introduce a new interface between the free space light and the WGMs of circular resonators. This interface is formed by directly depositing nano-scatterers or nanoparticles onto the circular WGMR. We show that each of the nanoparticle deposited on the resonator surface effectively acts as a nanocoupler to couple free space light into WGMs without additional bulk optical components and precise alignment processes. Further, we demonstrate lasing in an Ytterbium ($Yb^{3+}$) doped silica microtoroid. Cavity-enhanced Rayleigh scattering lies at the heart of our nano-scale interface between the micro-scale WGMR and the free-space light field[27]. The hybrid microresonator-nanoparticle system here enables the collection of a large fraction of the scattered light into the cavity mode via Purcell enhancement, and has the ability to harvest even weak light fields. This nanocouplers scheme brings together and relies on four fundamental observations. First, coupling of an emitter to a cavity mode enhances its spontaneous emission rate by increasing the local density of modes, implying that the emitter will emit mostly into the cavity modes and with much faster rate than in vacuum. This enhancement is proportional to Q/V and is known as Purcell enhancement factor. Second, a subwavelength particle (i.e., the nanocoupler) can be treated as an oscillating dipole, with the dipole moment induced by the electric field of the incident light, radiating into the surrounding (i.e., Rayleigh scattering). For the resonator, there is no difference between the light coming from an emitter placed its proximity and the light coming via scattering from a nanoparticle illuminated by a free-space incident light. Thus, Purcell enhancement should take place leading to collection of the weak scattered-light into the cavity WGM. Third, when a nanoparticle is placed close to a resonator and interact with the evanescent field of the resonator, light scattering back into the WGM and also to the free-space reservoir modes takes place. Here, Purcell effect manifests itself again by enhancing the coupling of the scattered light back into the degenerate WGMs (i.e., over 95% of the scattered light is coupled back[28, 29]). Fourth, nanoscatterers on the resonator break its rotational symmetry thus open a channel for coupling light in and out of WGMs[30, 31]. Therefore, the proposed nanocoupler should provide an efficient route for free-space coupling of light to and from WGMs. An illustration of the concept is given in Fig. 1

We have developed a theoretical model (Supplementary Note 1) and performed extensive numerical simulations (Supplementary Note 2) to quantify the interaction between free-space light and a WGMR with and without perturbing nanoparticles of spherical shape. Figure 1**c** clearly shows that in the absence of nanoparticles, free-space light does not couple into WGMs while the presence of a single nanoparticle opens up a channel, which interfaces the WGMs inside the resonator with the outside optical modes, including the free-space incident light and the reservoir modes into which the WGMs dissipate.

## Results

The setup used in the experiments is depicted in Fig. 1**b**. It consists of a tunable external cavity laser and a fiber lens as the free-space light source, a fiber-taper coupler to extract the light, which is coupled into the microtoroid from free-space via the nanocouplers, out from the WGM, and a nanoparticle delivery system to deposit the nanoparticles onto the resonator (see Supplementary Fig. 1 and Method 1 for details). The light extracted from the WGM through the fiber-coupler is evaluated using optical spectrum analyzer (OSA) and photodiodes. Fiber-taper coupler enables efficient and tunable out-coupling of the light from the microtoroid for accurate evaluation of the proposed nanocoupler. To evaluate the performance of the nanocoupler, we performed three sets of experiments using this scheme.

In the first set of experiments, we investigated the effect of nanocoupler-resonator and resonator-taper coupling strengths, quantified by $2\Gamma$ and $\kappa_1$, respectively (see Supplementary Note 1), on the intracavity power using a free space light whose spot size was larger than the area of the resonator so that all the nanocouplers were illuminated. In this way, we investigated the collective response of all possible coupling channels (nanocouplers). We first deposited nanoparticles until enough power can be transmitted from free space into the resonator. We specifically chose a large resonator having a large V to prevent or reduce the possibility of observable mode splitting (i.e., amount of mode splitting scales inversely with V). Using a nanopositioning system, we changed $\kappa_1$ by tuning the taper-resonator distance and monitored the out-coupled light from fiber taper as the wavelength of the free-space light is scanned (Fig. 2**a**). Since the WGM resonator supports two counter-propagating modes (clockwise, CW and counterclockwise, CCW) at the same resonance frequency, the frees pace light is coupled into both CW and CCW modes. We observed clear resonance peaks in both the forward (PD1) and the backward (PD2) directions implying the coupling of free-space light into WGM via the nanoparticle based interface (nanocoupler). The extracted light intensity changed as the distance between the resonator and the fiber taper (i.e., air gap) was changed (Fig. 2**a, b**). As shown in Fig. 2a, in the deep-under-coupling regime ($\kappa_1 \ll \kappa_0+2\Gamma$), the sum of the intrinsic resonator loss $\kappa_0$ and the loss induced by the nanocouplers $2\Gamma$ dominated the fiber-coupling loss $\kappa_1$; thus, only a small amount of light was extracted from WGM. Decreasing the air gap brought the system closer to the critical coupling ($\kappa_1 = \kappa_0+2\Gamma$), which was accompanied by an increase in the extracted light power at the resonance wavelength; the extracted power reached its maximum value at the critical coupling condition. Further decrease of the air gap moved the system to over-coupling regime (i.e., fiber-coupling loss dominates other losses $\kappa_1 \gg \kappa_0 + 2\Gamma$) and led to reduction in the extracted peak power and to an increase in the linewidth of the resonance peak due to increased loss. We estimated the $Q$ as $10^6$ at the critical coupling point, and $\sim 2 \times 10^6$ at the deep-under-coupling regime, which is consistent with the definition of critical coupling condition (Fig. 2**b**).

Increasing the number of nanoparticles (channels, nanocouplers) increased the number of coupling channels between the WGM and its environment (i.e., increasing $2\Gamma$) and thus led to broader resonance linewidth (i.e., $2\Gamma + \kappa_0 + \kappa_1$) and to more light coupled from free space to WGMs at resonance (Fig. 2**c,d**). For each measurement depicted in Fig. 2**c,d**, we fixed the number of nanoparticles and optimized $\kappa_1$ such that $\kappa_1 = \kappa_0 + 2\Gamma$ was satisfied. Thus, power extracted from the WGM via the taper was maximized. In this case, the maximum intracavity power was achieved at the critical coupling of nanocoupler-resonator system quantified by $2\Gamma = \kappa_0$ (Fig. 2**d**). Further increase of the number of nanocouplers (larger $2\Gamma$) shifted the system beyond this critical point and lowered the intracavity power by increased particle-induced dissipation. These results (Fig. 2) are similar to what has been observed for an add-drop filter configuration where a resonator is coupled to two fiber taper couplers simultaneously[32]. The experimentally observed dependence of extracted power and the intracavity power on $\kappa_1$, $\kappa_0$ and $2\Gamma$ agrees well with theoretical model (see Supplementary Note 1).

In the second set of experiments, we started with a resonator-nanocouplers system with observable mode splitting in the transmission spectra (Fig. 3**b**) and investigated the response of local channels (nanocouplers) by changing the position of a tightly focused free-space beam spot along the equator of a microtoroid. In this way, we opened only the local channels within the small beamspot size for coupling free space light into WGMs, and recorded the extracted light in both forward (PD1) and backward (PD2) directions as a function of the beamspot position. In such a case when mode splitting exists, the light inside the resonator is expressed as two orthogonal standing wave modes formed by the superpositions of the CW and CCW modes[30, 31, 33]; the first standing wave mode SWM1 is expressed as $a_{swm1} = (a_{CW} + a_{CCW})/\sqrt{2}$ and the second standing wave mode SWM2 is expressed as $a_{swm2} = (a_{CW} - a_{CCW})/\sqrt{2}$. The spatial distribution of these modes are $\pi/2$-phase shifted from each other with the phase describing the spatial distance between the nodes of the SWMs, and the distance between two adjacent nodes of a SWM corresponding to $\pi$. As we scanned the beamspot along the resonator surface, we observed distinct spectra at different positions due to the fact that at each position different sets of local nanocouplers were excited and consequently different local channels were opened. Interestingly, we observed mode splitting in both the forward and the backward transmission spectra. While some sets of local channels coupled light strongly into the SWM1, the other sets of local channels coupled light strongly into the SWM2. This reveals that the nanocouplers channel more free space light into the modes that has stronger spatial overlaps with the nanocouplers. If the excited couplers are closer to the node of a mode, either no light or only a small amount of light can be coupled to that mode. These features are clearly seen in Fig. 3**b**, where split modes show different heights at different probe positions. The discrepancy in the intensity of the two modes is attributed to the different coupling channels and the variation (phase and amplitude) in the light coupled to each of these channels. We also observed that the intensities of the light coupled out in the forward and backward directions were not the same (Fig. 3). This is because the placement of nanoparticles was not symmetric with respect to forward and backward directions, which gave different phase for the forward and backward light. In this case the CW and CCW light inside the resonator were no longer the same.

In the third and final set of experiments, we demonstrated that the free-space-to-WGM coupling efficiency of the proposed nanocoupler scheme was sufficient to obtain WGM microlasers (Fig.4). Using free-space light from a tunable laser in the 980nm band, we observed WGM lasing in the 1050 nm band from an $Yb^{3+}$-doped silica microtoroid[34, 35]. Lasing started when sufficiently high pump power was built-up in the microcavity (Fig. 4**c**). The lasing threshold depends on the spatial overlap between the lasing and the pump mode as well as the spectral overlap between WGMs and the absorption and emission bands of $Yb^{3+}$ ions. We observed not only single mode lasing but also multimode lasing within the emission band of $Yb^{3+}$ ions (Fig. 4**b**). In multimode lasing, mode competition among many modes in the WGMRs may affect the lasing threshold.

## Discussion

In summary, we have introduced a simple yet elegant and efficient nanocoupler scheme to couple free-space light into WGMRs. The nanocoupler is based on nano-scatterers deposited onto the WGMR. We have observed enhanced light coupling into the WGM through the channels opened by this nanocoupler scheme, and demonstrated WGM lasers using free-space pumping of the resonator via this nanocoupler. One can further enhance the coupling efficiency by placing the nanoscatterers in well-defined locations on the resonator with equal distances between them to facilitate constructive interference. We believe that this study lays the foundation for future on-chip solar/non-coherent light pumped microlasers, and will greatly facilitate miniaturized WGM resonator applications without complicated coupling optics, and free from chaotic behavior often observed in deformed cavities. Also significant is the inherent local enhancement near the nanoparticles which can dually serve for considerably enhancing sensing[16].

**Acknowledgments** This work was partially supported by Army Research Office grant No. W911NF-12-1-0026.

**Author Information** The authors declare no competing financial interests. Correspondence and requests for materials should be addressed to J.Z. (jzhu@seas.wustl.edu), S.K.O. (ozdemir@ese.wustl.edu) and L.Y. (yang@seas.wustl.edu).

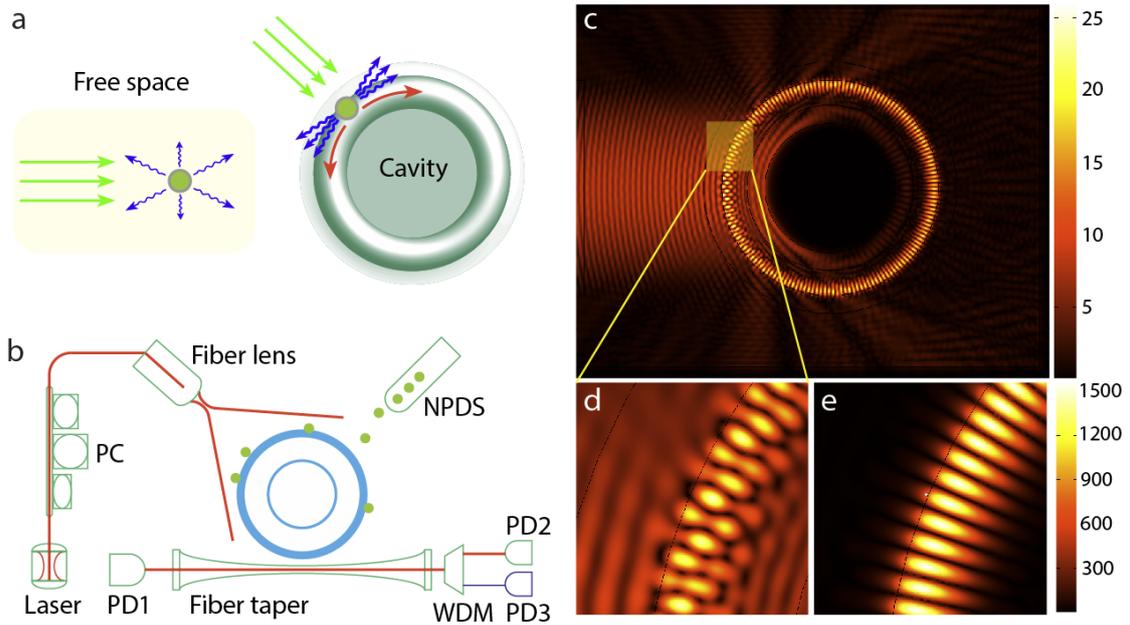

**Figure 1 | Coupling of free space light into whispering gallery modes via nano-scatterers. a,** Cavity enhanced light collection. A nanoscale structure usually scatters light in all directions; when a nano-scatterer is placed close to a cavity, most of the scattered light is collected into the cavity mode due to Purcell enhancement. **b,** Simplified scheme used for characterizing the performance of the nanocoupler. Light from a tunable laser is sent through a fiber lens and incident onto the resonator. The distance between the fiber lens and cavity is tunable. A fiber taper is used to monitor the light field inside the cavity. PC: polarization controller. WDM: wavelength division multiplexer. NPDS: nanoparticle delivery system. **c,** Finite element simulation shows free space light cannot couple into a WGM resonator by direct illumination. **d,** Magnified view of the WGM area in **c**. **e,** With a nanoparticle on the surface of resonator, WGM is efficiently excited from free space.

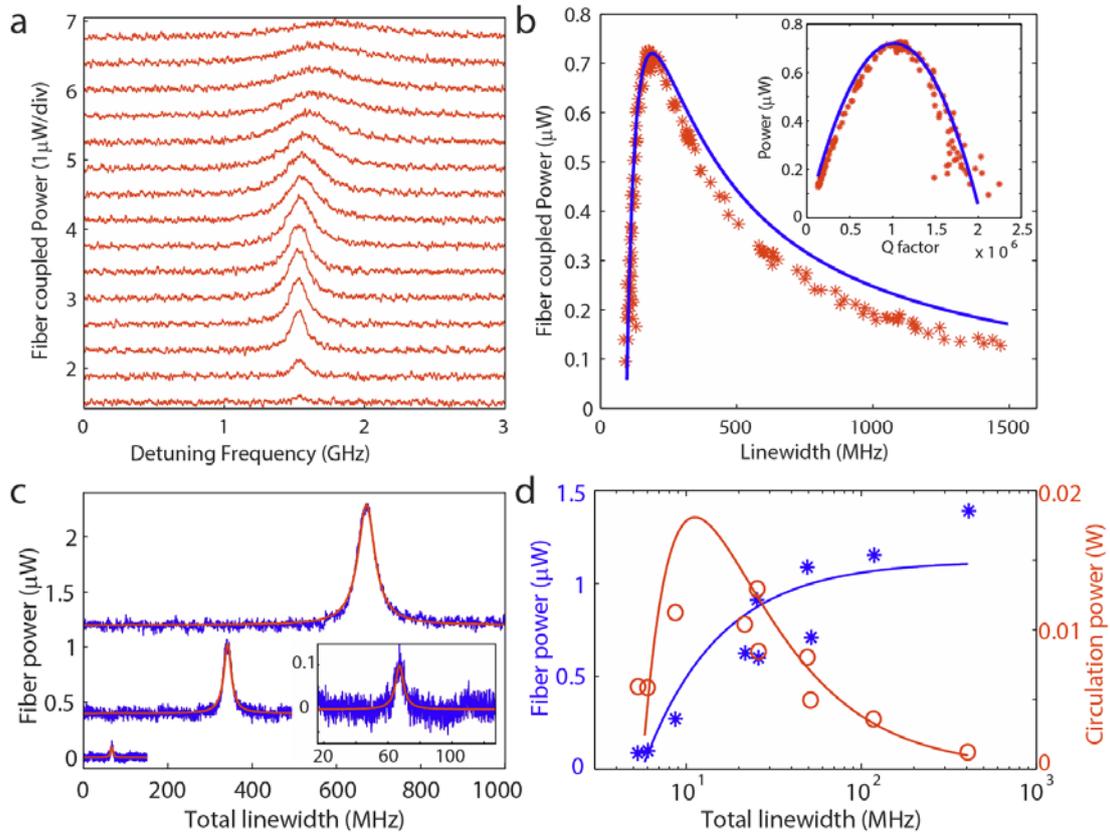

**Figure 2| Loading curves of nanocoupler-resonator-taper system. a,** Spectra of a WGMR, with input light coupled from free space into the resonator via nanocouplers, obtained at different fiber-taper coupling strength quantified by $\kappa_1$. Spectra are vertically shifted for clarity. From bottom to top: the air gap between the fiber taper and the resonator decreases ($\kappa_1$ increases). **b,** Power coupled out from the WGMR by the fiber taper versus the total linewidth of the nanocoupler-resonator-taper system when $\kappa_1$ is increased. Inset shows the power out-coupled using the fiber taper versus total Q factor. **c,** Mode spectra with increasing number of particles deposited on the microtoroid (increasing $2\Gamma$). Spectra are vertically shifted for clarity. From bottom to top: particle number increases. For each spectrum, fiber taper coupling is optimized to obtain the maximum on-resonance power ($\kappa_1 = \kappa_0 + 2\Gamma$). **d,** Blue asterix: power coupled from WGM using the fiber taper versus total linewidth of the system when more particles are deposited. Red circles: Calculated intra-cavity power versus total linewidth. Solid curves are fitting functions obtained from the theoretical model.

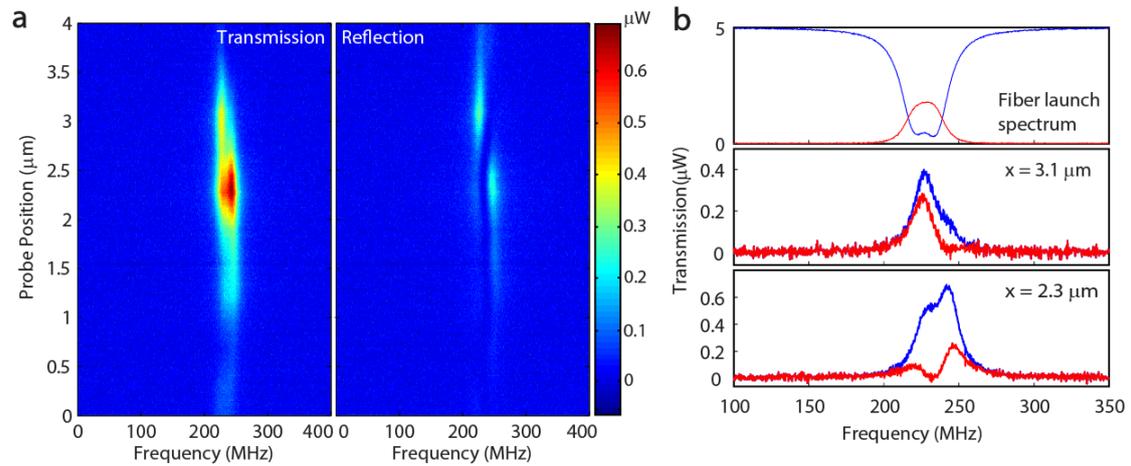

**Figure 3 | Mode spectra of a WGMR withincident beam spot at different positions. a,** Spectrograms of light coupled out of the WGMR by a fiber taper in the forward and backward directions when a tightly focused free-space beam spot was scanned along the equator of the microtorid with nanocouplers. The size of beam spot was less than 5 μm. **b,** Light coupled out from the microtoroid via a fiber taper when the same fiber taper was used to couple light into the microtoroid (top panel) and when free-space light is coupled into the microtoroid via the nanocouplers within the beam spot size (middle and bottom panels). The transmission spectra in the middle and bottom panels correspond to the probe positions of 3.1 μm and 2.3 μm in **a**. Different probe positions yield distinctively different spectra. Blue and red curves denote the transmission spectra in the forward and backward directions, respectively.

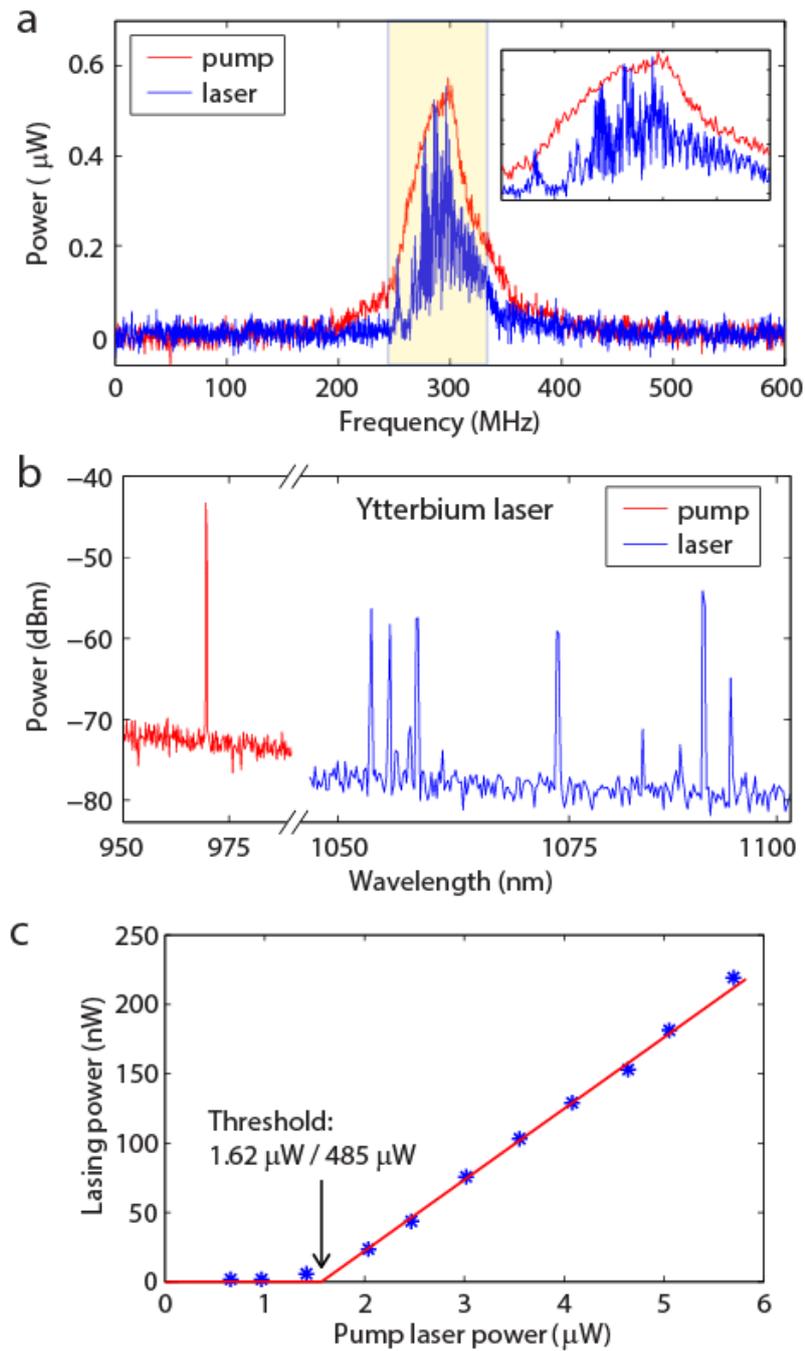

**Figure 4 | Whispering gallery mode microlasers by free space excitation via nanocouplers on a microtoroid resonator. a**, Lasing from an Ytterbium doped silica microtoroid achieved by free space excitation. The frequency of the free space pump at 975 nm band was scanned across 600 MHz range around a resonance. The red and blue spectra were obtained by out-coupling the

field in the microtoroid using a fiber-taper coupler, and they respectively correspond to the pump in the 975 nm band and the generated ytterbium lasing in the 1050 nm band. The diameter of the fiber taper coupler was optimized for maximal out-coupling of 1050 nm band light from the microtoroid. **b**, Pump and lasing spectrum of the Ytterbium laser shown in **a**. **c**, Relationship between lasing power and pump power of a Ytterbium laser. The threshold pump power for the lasing was measured at two different points, and was found to be 1.62 µW when the power was measured at the output of the fiber taper and 485 µW when it was measured at the input end of the fiber lens.

# Supplementary Information -

# Nanocouplers for free-space excitation of whispering-gallery microlasers

JiangangZhu[1*], Şahin K. Özdemir[1*], Huzeyfe Yilmaz[1], Bo Peng[1], Mark Dong[2], Matthew Tomes[2], Tal Carmon[2], and Lan Yang[1*]

[1]Department of Electrical and Systems Engineering, Washington University, St. Louis, MO 63130, USA

[2]Department of Electrical and Computer Engineering, University of Michigan, Ann Arbor, MI 48109, USA

*Correspondence to: yang@ese.wustl.edu, ozdemir@ese.wustl.edu, jzhu@seas.wustl.edu

**Supplementary Figures**

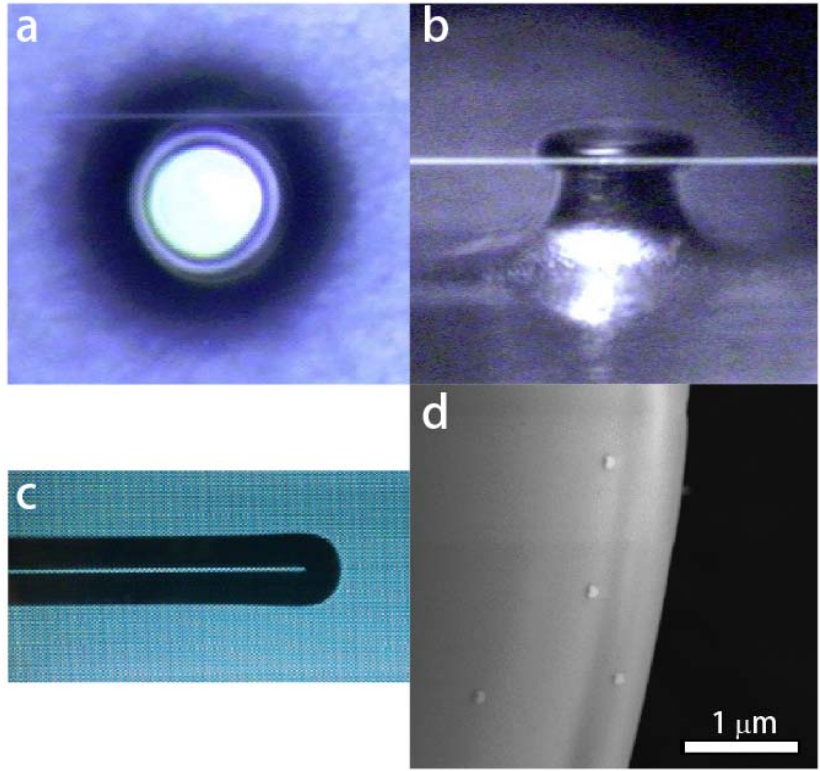

**Supplementary Figure S1 | Experimental images. a,** Image of a microtoroid with diameter of about 80 μm coupled with a tapered fiber, top view. **b,** side view of **a. c,** Optical image of the fiber lens made from a single mode fiber (SMF-28e). **d,** Polystyrene nanoparticles (radii 50 nm) deposited on the surface of a microtoroid.

# Supplementary Method 1

*Fabrication and Experiment setup*

To study the free space light coupling using nanocouplers, we fabricated passive microtoroids with major diameters of 60-150 μm using commercially available silicon wafers with 2 μm oxide layer. Active microtoroids were fabricated using sol-gel method[16], either from an Ytterbium ion doped ($Yb^{3+}$ concentration: $2 \times 10^{18}/cm^3$ to $2 \times 10^{19}/cm^3$) sol-gel silica layer (3 layer coating, 1.5 μm total thickness) on silicon or by coating a reflowed microtoroid with rare-earth ion doped sol-gel film followed by $CO_2$ laser annealing. The details of fabrication process can be found in reference 16 in the main text. Fabricated microtoroids had optical Q factors of $10^7$-$10^8$ for passive resonators and above $10^7$ for active ones. We have also fabricated passive microspheres by melting the end of tapered fibers using fusion splicer (Ericsson FSU 995FA) or $CO_2$ laser pulses. Active microspheres were made by dip-coating passive ones into the sol-gel solution containing $Yb^{3+}$ (ion concentration $2 \times 10^{18}/cm^3$ to $2 \times 10^{19}/cm^3$) followed by high temperature annealing. The resulting passive and active microspheres had Q factors between $10^7$ and $10^8$.

Tapered fibers[6,7] used to extract light from microresonators were fabricated by pulling single mode fibers on a hydrogen flame. In order to couple the pump light (free space light coupled into the resonator via the proposed nanocouplers) and the generated laser light (from the rare-earth ions doped into the silica resonators) out from the WGMs of the resonators, we fabricated fiber tapers of varying sizes and selected the ones which achieved optimal or close-to-optimal coupling efficiency. It is worth noting that, due to the wavelength difference between the pump and lasing modes, optimal coupling for the pump (lasing) mode doesn't ensure optimal condition for the lasing (pump) mode. To send laser light from free space to the microresonator, we fabricated a simple lensed fiber by cleaving a single mode fiber followed by partially reflowing the end using a fusion splicer (Ericsson FSU 995FA). The resulting fiber lenses (Fig. S1c) have focusing distances in the range of 20-100 μm and beam spot diameters around 5-10 μm in 1550 nm wavelength band.

Nanoparticles of known sizes and materials were delivered onto the surface of a resonator by using a system similar to that described in Ref. 4 and 5 in the main text. We used polystyrene nanospheres with radii of 20-75 nm (Thermo Scientific 3000 Series Nanosphere™ Size Standards) as nanocouplers in our experiments. Nanospheres were dispersed in water and sent through the aerosol particle classification system and finally carried out by ultra-high purity nitrogen and blown onto a resonator using a glass nozzle with end diameter of about 100 μm. The nozzle allows targeting specific microtoroid for deposition. The distribution of particles on resonator is governed by the nozzle angle and position. Deposited nanoparticles were verified using SEM (Fig. S1d).

We probed the WGMs of the resonators using tunable diode lasers (New focus Velocity, TLB-6300-LN controller) in 1550 nm or 980 nm bands. The tunable laser was modulated with a triangle wave signal to fine scan its wavelength across 0.08 nm range, and then was coarsely tuned to match with the resonance modes. The out-coupled light from the fiber taper was monitored with photodetectors (New Focus 1811-FC Photoreceiver, 900-1700 nm, 125 MHz) and data was acquired to a computer through oscilloscope in real time.

The fiber lens was mounted on a 3D translational stage and pointed at the microtoroid at an angle of about 15 degrees with respect to the plane of the silicon chip. We adjusted the distance between the lens and the resonators using a 3D translational stage to allow maximal coverage of the microtoroid, or, a minimal beam spot on the microtoroid.

## Supplementary Note 1

*Free space coupling through a single nanoparticle*

For a WGM cavity with quality factor Q and effective circumference L, we have:

$$\frac{P_{cav}}{P_{fs}^{\Omega_{cav}}} = \frac{2Q\lambda}{\pi n L} \tag{S1}$$

where $P_{fs}^{\Omega_{cav}}$ denotes the power scattered into cavity mode in both clockwise (CW) and counter-clockwise (CCW) directions under free-space condition and $P_{cav}$ represents the power scattered into the cavity mode, which should be equal to the power dissipated by the cavity. Due to *Purcell* enhancement[27-29], and taking into account the normalized field distribution $f(\vec{r})$ at the particle location $\vec{r}$, we find

$$\frac{P_{cav}}{P_{dip} \cdot f^2(\vec{r})} = P = \frac{3Q\lambda_n^3}{4\pi^2 V} \tag{S2}$$

where $P_{dip}$ designates total power scattered into the dipole mode in free space, $\lambda_n$ denotes the effective wavelength of the WGM mode and V is the mode volume defined as:

$$V = \frac{\int_V \varepsilon(\vec{r})|E(\vec{r})|^2 \, dV}{\max(\varepsilon(\vec{r})|E(\vec{r})|^2)} \tag{S3}$$

Thus, we have:

$$\frac{P_{fs}^{\Omega_{cav}}}{P_{dip}} = \frac{P_{fs}^{\Omega_{cav}}}{P_{cav}} \frac{P_{cav}}{P_{dip}} = \frac{\pi n L}{2Q\lambda} \frac{3Q\lambda_n^3}{4\pi^2 V} f^2(\vec{r}) = \frac{3L\lambda_n^2}{8\pi V} f^2(\vec{r}) \tag{S4}$$

Since $P_{dip}$ can be evaluated by multiplying the input intensity $I_{in}$ (or $E_{in}^2$) and the particle cross-section $\sigma_s$, we can calculate $P_{fs}^{\Omega_{cav}}$ using:

$$P_{fs}^{\Omega_{cav}} = \frac{3L\lambda_n^2}{8\pi V} f^2(\vec{r}) \cdot I_{in} \sigma_s \tag{S5}$$

This relation can also be obtained similarly by evaluating the overlap integral between dipole radiation mode and a Hermite-Gaussian cavity mode in a mirror cavity[27]. Next, we can write the cavity rate equations with one particle of radius $r$:

$$\frac{dE_{CW}}{dt} = -(i(\Delta\omega+g)+\frac{\kappa_0+\kappa_1+\Gamma}{2})E_{CW} - (ig+\frac{\Gamma}{2})E_{CCW} - i\sqrt{\frac{3L\lambda_n^2}{16\pi V}f^2(\vec{r})\cdot I_{in}\sigma_s}$$

$$\frac{dE_{CCW}}{dt} = -(i(\Delta\omega+g)+\frac{\kappa_0+\kappa_1+\Gamma}{2})E_{CCW} - (ig+\frac{\Gamma}{2})E_{CW} - i\sqrt{\frac{3L\lambda_n^2}{16\pi V}f^2(\vec{r})\cdot I_{in}\sigma_s}$$

$$a_{fiber} = iE_{CW}\sqrt{\tau_0\kappa_1} \qquad (S6)$$

where $\Gamma = \frac{\alpha^2 f^2(\vec{r})\omega_c^4}{6\pi\upsilon^3 V}$, $g = -\frac{\alpha f^2(\vec{r})\omega_c}{2V}$, and $\alpha = 4\pi r^3(n_p^2-1)/(n_p^2+2)$. Define $E_{sym} = (E_{CW}+E_{CCW})/\sqrt{2}$ as the symmetric eigenmode (the asymmetric eigenmode doesn't receive any input power because the particle is situated at its node). From the above differential equations, we find the rate equation for the symmetric mode as:

$$\frac{dE_{sym}}{dt} = -(i(\Delta\omega+2g)+\frac{\kappa_0+\kappa_1+2\Gamma}{2})E_{sym} - i\sqrt{\frac{3L\lambda_n^2}{8\pi V}f^2(\text{r})\cdot I_{in}\sigma_s} \qquad (S7)$$

At resonance condition $\Delta\omega+2g=0$, we can find the steady state solution of $E_{sym}$:

$$|E_{sym}| = \frac{\sqrt{\frac{3L\lambda_n^2}{2\pi V}f^2(\text{r})\cdot I_{in}\sigma_s}}{\kappa_0+\kappa_1+2\Gamma} = \frac{\sqrt{\frac{3L\lambda_n^2}{2\pi V}f^2(\text{r})\cdot I_{in}\frac{8\pi^3}{3}\frac{1}{\lambda^4}\alpha^2}}{\kappa_0+\kappa_1+\frac{\alpha^2 f^2(\text{r})\omega_c^4}{3\pi\upsilon^3 V}} = \frac{\sqrt{\frac{4\pi^2 L}{V\lambda_n^2}f^2(\text{r})\cdot I_{in}}}{(\frac{\kappa_0+\kappa_1}{\sqrt{2\Gamma}}+\sqrt{2\Gamma})\frac{f(\text{r})\omega_c^2}{\sqrt{3\pi\upsilon^3 V}}}$$

(S8)

Here we used $\sigma_s = 8\pi^3\alpha^2/3\lambda_n^4$ and $\Gamma = \alpha^2 f^2(\vec{r})\omega_c^4/(6\pi\upsilon^3 V)$. If $\alpha$ or $\Gamma$ is the only variable, when $2\Gamma = \kappa_0+\kappa_1$, the intensity of the symmetric mode $|E_{sym}|^2$ is maximized (Cauchy–Schwarz inequality). This means that for a single scatterer when the particle induced loss equals to the sum of the all other cavity losses, the intra-cavity power is maximized. Finite-element simulations in the next section verifies this conclusion. It is worth noting that this conclusion is also correct for multiple particle case, as suggested in Fig. 2 in the main text.

On the other hand, for the fiber taper output, we have:

$$|a_{fiber}| = |iE_{CW}\sqrt{\tau_0 \kappa_1}| = \sqrt{\tau_0 \frac{3L\lambda_n^2}{4\pi V} f^2(\mathrm{r}) \cdot I_{in}\sigma_s} / (\frac{\kappa_0 + 2\Gamma}{\sqrt{\kappa_1}} + \sqrt{\kappa_1}) \tag{S9}$$

If fiber taper coupling $\kappa_1$ is the only variable parameter, then we find that setting it as $\kappa_1 = \kappa_0 + 2\Gamma$ (critical coupling condition) maximizes the output of fiber taper (Fig. S2). This conclusion is consistent with experimental results in Fig. 2.

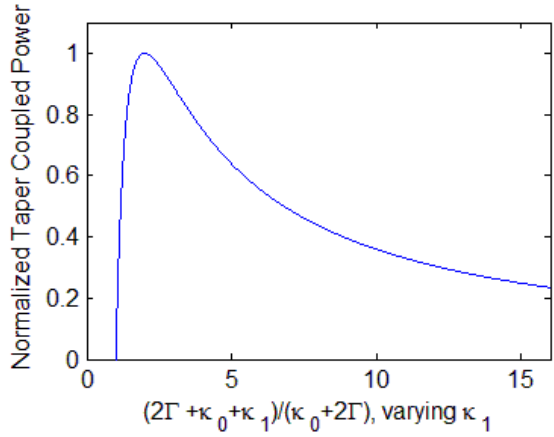

**Supplementary Figure S2 |** Fiber taper coupled power $|a_{fiber}|^2$ versus normalized cavity loss (normalized to $\kappa_0 + 2\Gamma$) with varying $\kappa_1$

*Free space coupling through multiple nanoparticles*

When there are multiple scattering centers on the resonator, the scattering light from each scatterer needs to be taken into consideration. Assuming there are N particles, and the light field incident on particle $n$ is $E_{in}^n e^{i\varphi_n}$, the rate equations can be expressed as:

$$\frac{dE_{CW}}{dt} = -(i(\Delta\omega + g_{sym} + g_{asym}) + \frac{\kappa_0 + \kappa_1 + \Gamma_{sym} + \Gamma_{asym}}{2})E_{CW} - (i(g_{sym} - g_{asym}) + \frac{\Gamma_{sym} - \Gamma_{asym}}{2})E_{CCW} - i\sum_{n=1}^{N} E_{CWin}^n$$

$$\frac{dE_{CCW}}{dt} = -(i(\Delta\omega + g_{sym} + g_{asym}) + \frac{\kappa_0 + \kappa_1 + \Gamma_{sym} + \Gamma_{asym}}{2})E_{CCW} - (i(g_{sym} - g_{asym}) + \frac{\Gamma_{sym} - \Gamma_{asym}}{2})E_{CW} - i\sum_{n=1}^{N} E_{CCWin}^n$$

$$\tag{S10}$$

where $E_{CWin}^n = \sqrt{\frac{3L\lambda_n^2}{16\pi V}} f_n^2(\vec{r}) \cdot \sigma_s^{\ n} E_{in}^n e^{i(\varphi_n+\beta_n-\phi_N)}$, $E_{CCWin}^n = \sqrt{\frac{3L\lambda_n^2}{16\pi V}} f_n^2(\vec{r}) \cdot \sigma_s^{\ n} E_{in}^n e^{i(\varphi_n-\beta_n+\phi_N)}$,

$\beta_n \in [0, 2\pi)$ is the spatial phase position of the *n*-th particle[31] assuming that the distance between the nodes of standing wave mode (SWM) corresponds to $2\pi$, $\varphi_n \in [0, 2\pi)$ is the phase of light incident on the *n*th particle and $\phi_n \in [0, 2\pi)$ is the phase position of the antinode of the symmetric SWM[30,31,33]. The scattering and damping terms are defined as[30,31]:

$$g_{sym} = \sum_{n=1}^{N} g_n \cos^2(\phi_N - \beta_n), g_{asym} = \sum_{n=1}^{N} g_n \sin^2(\phi_N - \beta_n)$$
$$\Gamma_{sym} = \sum_{n=1}^{N} \Gamma_n \cos^2(\phi_N - \beta_n), \Gamma_{asym} = \sum_{n=1}^{N} \Gamma_n \sin^2(\phi_N - \beta_n)$$
(S11)

where $g_n$ and $\Gamma_n$ are the scattering and damping coefficients of *n*th particle. The phase position of standing wave modes can be calculated by[31,33,36]:

$$\tan(2\phi_N) = \frac{\sum_{n=1}^{N} g_n \sin(2\beta_n)}{\sum_{n=1}^{N} g_n \cos(2\beta_n)}$$
(S12)

## Supplementary Note 2

*Finite-element Simulations*

To verify our theoretical model and experimental results, we performed finite-element simulations using Comsol Multiphysics. A 2D model is presented to demonstrate the fundamental principles of the nanocoupler scheme. We also calculate the coupling of light into a microtoroid via a nanosphere by using a full 3D model [37].

*Free space excitation with and without a single nanoparticle*

We built a 2D WGM resonator with diameter of 30 μm and refractive index of 1.45-0.00000001i, surrounded by air (refractive index of 1). The small imaginary part in the refractive index is introduced to simulate the absorption and scattering loss in experiments. The Q factor of the resonator in simulation is $7.375 \times 10^7$. The center of the resonator (outside WGM area) is set to have strong absorption (refractive index of 1-0.5i) to minimize lensing effect by the curvature of resonator-air interface. A free space light port with 30 μm width is placed to the left of the resonator.

First, eigenmode analysis was used to find the resonance frequency and Q factor of the system. Second, harmonic stationary analysis was conducted and free space light with the same frequency of the resonance was sent from left. Figure S3 shows the simulation result. In this case, the average electrical field strength in the resonator is calculated to be 1.219. Although WGM pattern exists in the resonator, the electrical field strength inside the resonator is comparable to that of the input light in free space. In other words, WGM is not efficiently excited. With efficient excitation we mean that there is a significant discrepancy between the electric field strength in the resonator and that in the surrounding, i.e., the former being much larger than the latter due to resonance-enhancement of the field inside the resonator.

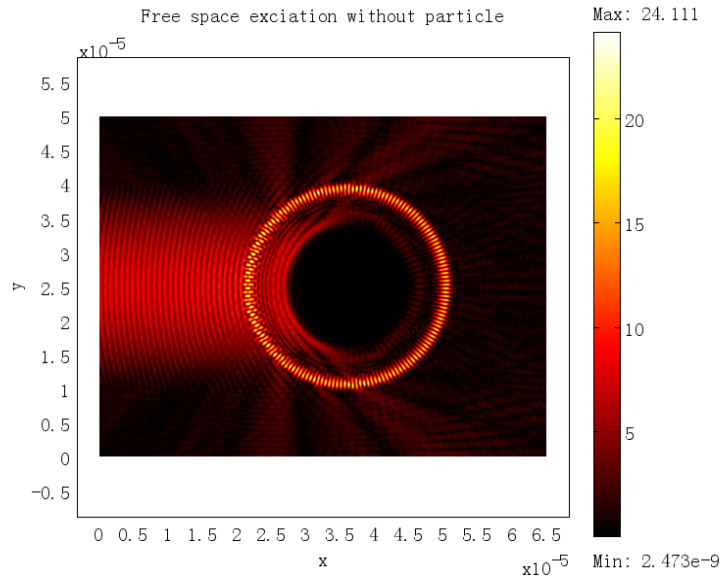

**Supplementary Figure S3** | Simulated electrical field distribution when free space light is illuminated on a bare resonator (without nanoparticles). WGM is not efficiently excited.

Next, we added a nanoparticle to the resonator surface and repeated the simulation. The nanoparticle had a radius of 20 nm and refractive index of 1.3. As expected the eigenfrequency of the resonator-particle system shows mode splitting. The Q factor of the symmetric mode (mode whose antinode overlaps with the nanoparticle) decreases to $2.942 \times 10^7$ and the Q factor of the asymmetric mode (mode whose node overlaps with the nanoparticle) is unchanged ($7.375 \times 10^7$) compared to the case without nanoparticle couplers. Then, harmonic stationary analysis was conducted and free space light with the same frequency of each split resonance mode was sent from left. Figure S4 shows the simulation results. For the symmetric mode, the light scattered from nanoparticle is efficiently collected into the mode due to optimal overlap between the mode and the particle, and the WGM is efficiently excited. On the other hand, the asymmetric mode receives very little scattered light due to minimal overlap between the mode and the nanoparticle. The average field strength in the resonator is calculated to be 85.75 and 1.286 for the symmetric and asymmetric mode, respectively. According to these numbers, a single nanoparticle increases the intra-cavity power by a factor of about 5000.

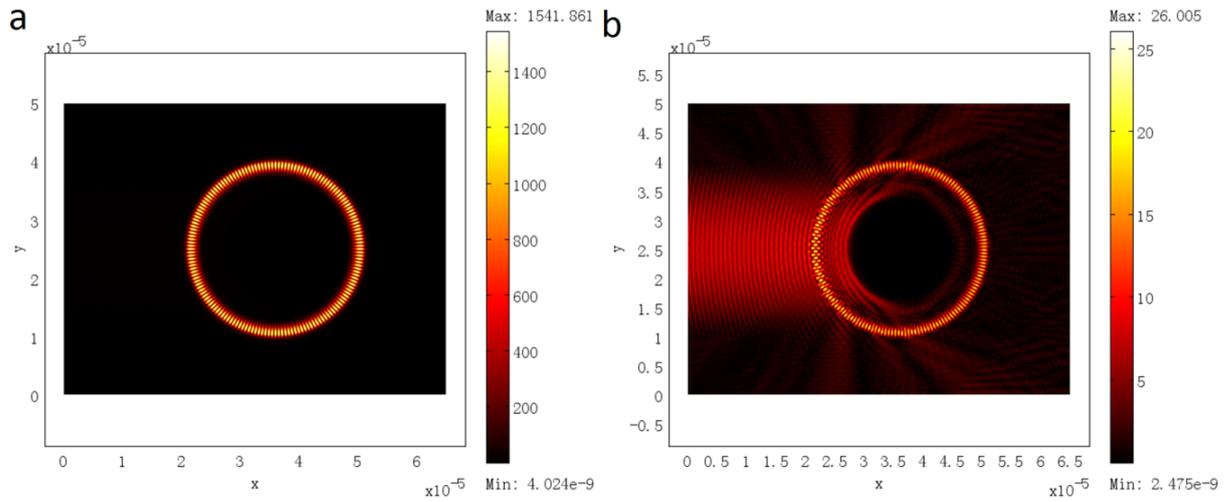

**Supplementary Figure S4 | Simulated electrical field distribution when free space light is illuminated on a resonator with a single nanoparticle. a,** Symmetric mode, which has maximal overlap with the nanoparticle, is efficiently excited. **b,** Asymmetric mode, which has minimal overlap with the nanoparticle, is not efficiently excited.

To obtain the frequency response of the resonator-particle system, we performed parametric analysis where the frequency of the free-space light was scanned and the average field strength in the resonator was calculated. Figure S5 shows the relation between the intra-cavity field and the frequency of the free-space light. The dominant resonance peak at the frequency of symmetric mode implies the efficiency of the proposed nanocoupler scheme to couple free-space light into the cavity and the resonance-enhancement of the intra-cavity field.

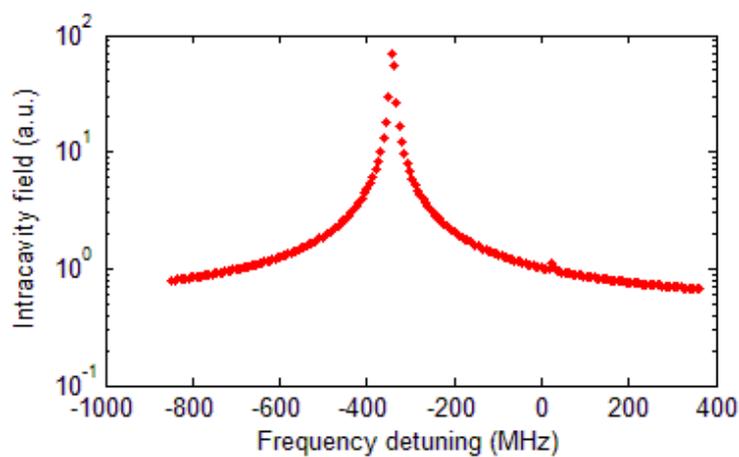

**Supplementary Figure S5** | Simulated intra-cavity field strength versus the frequency of the free-space light. A dominant resonance peak at the frequency of symmetric mode and a weak resonance peak at the frequency of asymmetric mode are clearly visible.

Next, we study the effect of nanoparticle polarizability on the intra-cavity field by varying the particle refractive index from 1 to 3. To compare with experimental results, we obtain the linewidth and Q factor of the symmetric mode when nanoparticle refractive index is increased. Figure S6 shows the relation between the intra-cavity power and the total linewidth or total Q factor. Simulation shows that when particle induced loss equals to the sum of all other cavity losses, the intra-cavity field is maximized, which agrees well with theory and our experimental results shown in the main text.

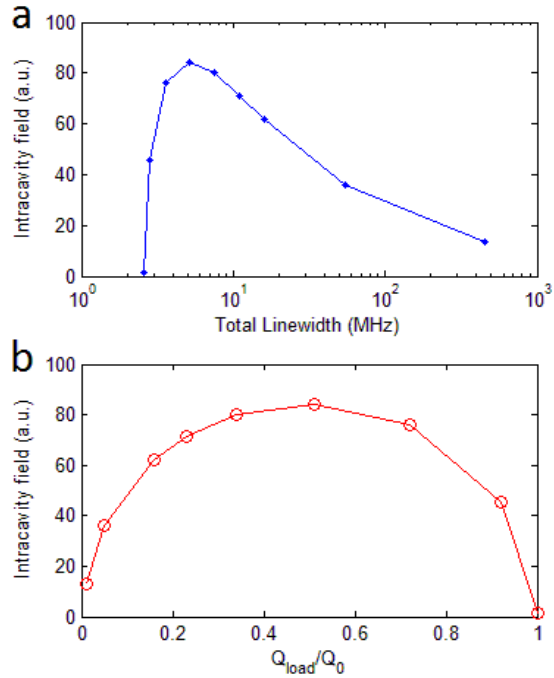

**Supplementary Figure S6 | Simulated intra-cavity field strength when nanoparticle induced loss is increased**. **a,** Intra-cavity field versus total linewidth. **b,** Intra-cavity field versus total Q factor (normalized by $Q_0$). The intra-cavity field strength reaches maximum when particle induced loss equals cavity intrinsic loss, or $Q_{load}= 2Q_0$.

*Free space excitation with two nanoparticles*

When there are multiple nanoparticles on the resonator, the eigenmodes (standing wave modes, SWMs) inside the resonator no longer situate their node or anti-node at the position of nanoparticles. The distributions of SWMs (split modes) are determined by the particle ensembles according to their relative spatial phase positions in the azimuthal direction on the resonator and their overlap $f(\vec{r})$ with WGM[30,31]. As discussed in Sec.2, intra-cavity power depends not only on the positions of the particles and their overlap with WGM, but also on the intensity and phase of the free-space incident light on each particle.

To study this phenomenon, we performed simulations on the simplest case of two particles on the resonator. Similar to the simulation model in the single particle case, we add another particle of the same size on the resonator; the two particles are arranged symmetrically (vertically mirrored) with respect to the axis of the resonator along the direction of the incident light (Fig. S7). The phase of the incident light on these two particles is kept the same, and the particles have the same distance to the resonator surface (same overlap $f(\vec{r})$ with WGM). The intensity of light incident on the first particle is decreased to about half of that on the second one by attenuating the incident light. Note that if the intensity of light incident on two particles is the same, only the symmetric mode can be excited because the system becomes symmetric around horizontal center line. The nanoparticles have radii of 20 nm and refractive index of 1.5-0.05i. The purpose of having an imaginary part for the refractive index of the nanoparticle is to increase the particle induced loss to compensate the difference in particle scattering properties between 2D and 3D cases. In 2D simulation, nanoparticles with the same diameter and refractive index usually give larger mode splitting but smaller particle- induced linewidth broadening compared to the 3D case.

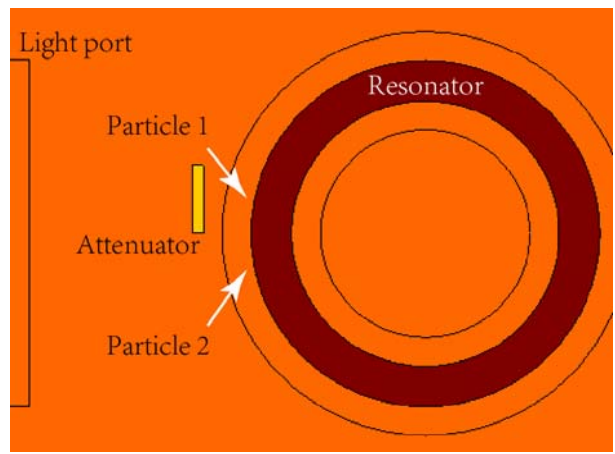

**Supplementary Figure S7 |Finite-element simulation model with two particles on a resonator.** Two nanoparticles (too small to be seen in the figure) are positioned symmetrically with respect to the horizontal axis of the resonator and light port. An attenuator with refractive index of 1-0.1i is placed in front of particle 1 to decrease the light intensity incident on the particle. The resonator has diameter of 30 μm and refractive index of 1.45-0.00000001i and the nanoparticles have radii of 20 nm and refractive index of 1.5-0.05i.

We then varied the relative spatial phase distance between the two particles from $\pi/8$ to $\pi$ at intervals of $\pi/8$, and calculated the relation between average intensity inside the resonator and wavelength using harmonic parametric simulation. Figure S8 shows the simulation results and compares them with theoretical calculations. Both theoretical and simulation results show mode crossing pattern which is due to the changes in the relative position of nanoparticles[31]. The simulation results for eigenmode position and field strength agree well the calculated results predicted by the theoretical model presented in Section 2. However there is slight difference between 2D simulation and theoretical calculation (based on 3D model): As seen in Fig. 8b, the mode with shorter wavelength has higher Q factor than the calculated results in Fig. 8a and thus it results in a higher intra-cavity power. This can be attributed to the difference in the particle-induced scattering loss in 2D and 3D models. This difference leads to higher Q factor in 2D simulations as the optical mode experiences less loss when the corresponding standing wave mode does not strongly overlap with the particles (Fig S8b). Consequently, a higher intra-cavity circulating power is predicted. Nonetheless, the simulations provide a verification of our theoretical model.

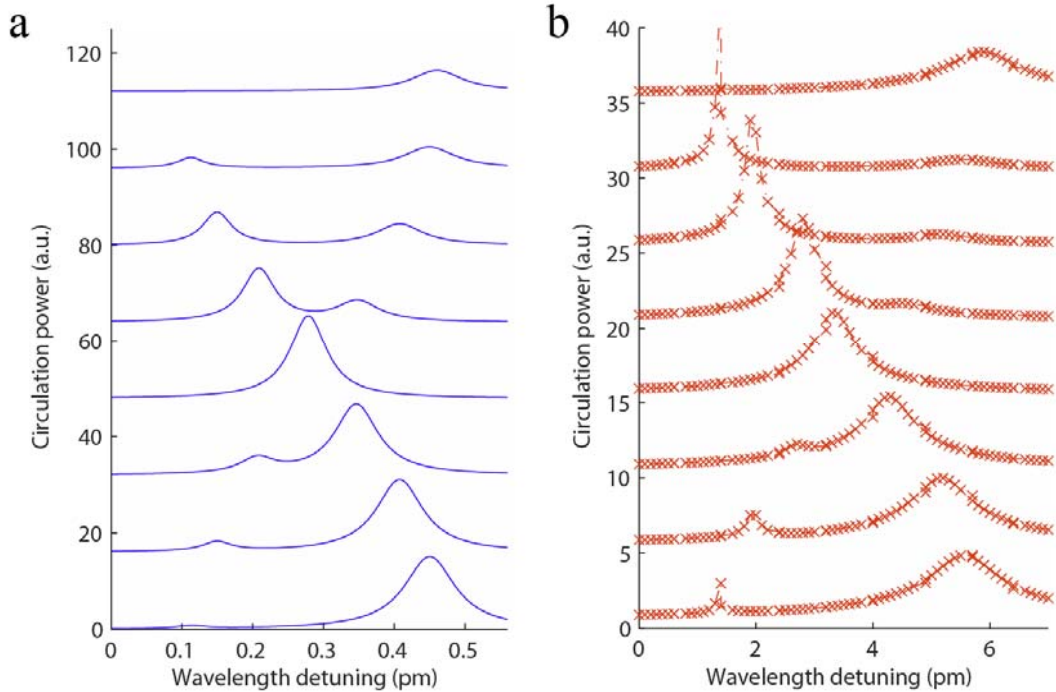

**Supplementary Figure S8 |Intracavity power with two nanoparticles on a resonator. The spectra are vertically shifted for clarity. From bottom to top: phase distance between the two particles is increased from $\pi/8$ to $\pi$ with $\pi/8$ interval. a,** Optical spectra of intra-cavity power calculated using the theoretical model. The nanoparticles have radii of 100 nm and refractive index of 1.59. Intrinsic Q factor of the resonator is $7\times10^7$. **b,** Optical spectra of intra-cavity power calculated using 2D finite-element simulations.

*3D model of a microtoroid-nanoparticle system*

We also calculate the coupling of light into a microtoroid via a nanoparticle using a fully 3D model [37]. The toroid is represented by a slice spanning 1.5 wavelengths along the circumference. As expected, with no nanosphere (Figure S9a, right) the field is minimally coupled to the surrounding [38]. On the contrary, when a nanosphere is attached to the toroid (Figure S9a, left) we see a quasi-spherical wavefront spreading around the nanosphere. In previous studies[28-29,39] this spherical wave represents light that is scattered out of the toroid by the nanoparticle. Here, on the contrary, this spherical wave represents light that is coming from the surrounding and coupled into the toroid. Both of the above interpretations are valid since this optical system is reciprocal [40]. Though the incoming field in our experiment is not identical with the one calculated, it was sufficiently similar to explain coupling of light into the toroid. Using the same numerical calculation [37], we check the coupling rate as a function of the nanoparticle size for various modes. This calculation shows a large difference between the coupling rates for various modes and allows, in principle, a critical coupling condition for the pump mode, while being under coupled for the laser mode in accordance with what is needed for low-threshold lasers.

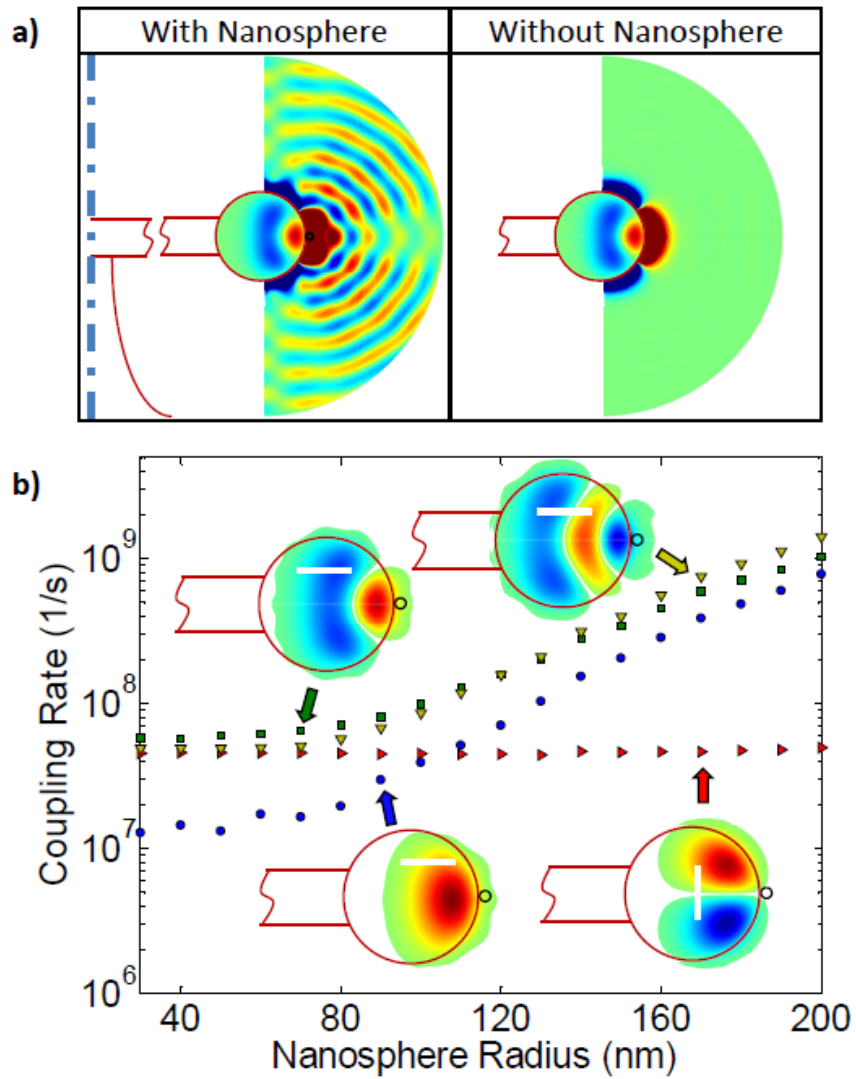

**Supplementary Figure S9 | Simulation results from a 3D model for light coupling from free space to WGMs by a nanoparticle**. **a**, The form of the electric field exhibits a quasi-spherical wave with origin at the nanoparticle. A test case with no nanoparticle

present is also shown for comparison. **b**, Parametric study of the coupling rate as a function of the nanoparticle size. The toroid major and minor diameters are 50 and 5 um. The microtoroid and the nanosphere are made of silica (n = 1.45) and polystyrene (n = 1.555) and the vacuum optical wavelength is near 1.55 μm. In **a**, the electric field outside the resonator has been scaled up by a factor of 100 to be visible, the nanoparticle radius is 200 nm and the optical polarization is horizontal. In **b**, the white lines represent polarization direction.